# Drivers of Electric Vehicle Adoption in Nigeria: An Extended UTAUT Framework Approach

Qasim Ajao [1]; Lanre Sadeeq [2]; and Oluwatobi Oluwaponmile Sadiq [3]

[1]Department of Electrical Engineering, Georgia Southern University, Statesboro, Georgia, USA. [2]Research Scholar and Subject Matter Expert, Microsoft Corporation, Ontario, Canada. [3]Department of Electrical Engineering, University of Lagos, Nigeria.

Corresponding author: qasim.ajao@ieee.org

**DOI**: https://doi.org/10.62154/ajesre.2024.016.010326

**Abstract**

Electric vehicles (EVs) represent a significant advancement in automotive technology, utilizing electricity as a power source in place of traditional fossil fuels and incorporating sophisticated navigation and autopilot systems. These vehicles align with multiple Sustainable Development Goals (SDGs) by offering a more environmentally sustainable alternative to internal combustion engine vehicles (ICEVs). Despite their potential, the adoption of EVs in developing nations such as Nigeria remains constrained. The Unified Theory of Acceptance and Use of Technology (UTAUT) framework is expanded in this study by including important enablers such as poor infrastructure, problems with affordability, and government support in the broader category of facilitating conditions. Additionally, it scrutinizes variables such as trust, performance expectations, social influences, and network externalities to identify the primary determinants influencing Nigerian consumers' propensity to adopt EVs. Results show that the percentage increase of H6 (facilitating conditions → behavioral intentions) compared to H5 (network externalities → behavioral intentions) is approximately 32.35%, indicating that traditional drivers significantly influence individuals' willingness to purchase EVs and are particularly strong factors in adoption decisions. The paper concludes with a discussion of these findings and proposes strategies for future research to further explore the barriers and drivers of EV adoption in Nigeria.

## Introduction

Battery electric vehicles (BEVs) and plug-in hybrid electric vehicles (PHEVs) are the two categories into which electric cars (EVs) fall. Battery electric vehicles (BEVs) depend on electricity as their primary power source, resulting in fewer emissions and lower operating expenses. Plug-in hybrid electric vehicles (PHEVs) use a combination of traditional internal combustion engines and electric propulsion systems, resulting in enhanced fuel efficiency and decreased emissions. As EV technology advances, it presents new opportunities for improving transportation, mitigating environmental effects, and alleviating financial constraints in the transportation sector. Nigeria, the leading oil-fired generator user in Africa, produces a significant 80% of its electricity from its abundant gas reservoirs. Therefore, even though there is an observable lean toward renewable energy sources in recent strategy discussions, natural gas continues to be pivotal for immediate power supply plans. EVs represent a major advancement in modern vehicle technology, offering a more environmentally friendly alternative to traditional gas-powered cars. (Aldhanhani, 2024). Primarily powered by electricity, these EVs boast advanced features like navigation sensors and autopilot capabilities. Often perceived as luxury items, EV combines cutting-edge technology with superior transportation, providing enduring economic benefits.

The United Nations has identified EVs as crucial for achieving their Sustainable Development Goals, which aim to create a cleaner and more sustainable future by 2030 (Gicha, 2024). Projections indicate that by 2030, imports will constitute 85% of the market, up from 70%, in Sub-Saharan Africa. However, most of these vehicles are second-hand imports, suggesting potential shifts in car ownership trends. Despite this increase in imports, Nigeria lags behind other regions in adopting smart vehicle technologies. Early adoption of EVs in countries like Mauritius, Seychelles, South Africa, and Rwanda within Sub-Saharan Africa has been observed. Forecasts suggest that Ethiopia, Kenya, Nigeria, Rwanda, and Uganda could potentially reach 4.0-5.0 million units in EV sales. The absence of government support, cost concerns, and insufficient infrastructure, particularly in the energy sector, present significant obstacles to the adoption of EVs in Nigeria. The lack of adequate infrastructure undermines the reliability and consistency of electricity distribution, which is crucial for the widespread acceptance of EVs. The erratic and unreliable energy supply diminishes the feasibility and appeal of EVs, which depend on a regular power supply for charging (Venkatesh, 2016) (Ajao, 2023).

This study uses an enhanced Unified Theory of Acceptance and Use of Technology (UTAUT) framework to examine the adoption of EVs in Nigeria. It considers various factors, including inadequate infrastructure, cost-effectiveness, government support, trust, performance expectations, social influence, and network externalities. This paper will examine how non-traditional factors within facilitating conditions influence the acceptance of EV technology in Nigeria. It will discuss key elements such as trust, social influence, and the enhanced UTAUT framework we developed. The study will also cover the research methodology, hypotheses, and data analysis, as well as the challenges and opportunities for EV adoption in Nigeria.

**Problem Statement**

Despite optimistic projections by industry leaders such as McKinsey and Shell, which estimate that EV sales in Nigeria and other Sub-Saharan African countries could exceed 1 million units within the next five years and potentially reach 5 million units by 2040, the actual rate of EV adoption in the region remains starkly limited. This discrepancy highlights a critical gap in understanding the unique factors that influence EV adoption in the region, where infrastructure, economic, and social conditions differ significantly from those in regions with higher EV penetration. Current models of technology acceptance, such as the UTAUT (Venkatesh, 2016), provide a framework for understanding user intentions and behavior. However, the specific context of EV adoption has not sufficiently adapted to these models. Performance expectancy, network externalities, effort expectancy, social influence, and trust are well-known concepts. However, they might not fully capture the complexities of EV adoption, where facilitating conditions and region-specific issues are very important.

This research addresses the urgent need to understand the factors driving or hindering EV adoption in Nigeria by enhancing the UTAUT framework to include additional determinants relevant to the region. It looks at how performance expectations, network externalities, effort expectations, social influence, trust, and tailored facilitating conditions—such as building reliable charging infrastructure, using renewable energy, and making helpful policies—affect decisions to adopt. By identifying and analyzing these factors, this study aims to fill the knowledge gap and provide actionable insights to accelerate EV adoption in the region, thereby aligning with global sustainability goals.

**Research Review**

The aim of researching why people choose to use EVs in Nigeria boils down to determining what drives them toward this innovative technology (Collett, 2021). We employ a useful tool known as the UTAUT introduced in 2003 (Venkatesh, 2016). This model has shown its utility in uncovering factors that affect adoption rates for smart technologies like EVs, where considerations extend beyond typical tech-adoption conundrums, such as balancing power grid development, regulating government incentives, or streamlining charging infrastructure, to also include abstract elements like trustworthiness and network externalities (Malima, 2023).

*UTAUT:*

The initial development of the UTAUT model revolved around four fundamental principles, or constructs: effort expectancy, facilitating conditions, social influences, and performance expectancy (Venkatesh, 2016). It unifies numerous theoretical perspectives and incorporates changeable influences through four balanced variables (experience, age, willingness to use, and gender), thereby enhancing its explanatory abilities. Despite its comprehensiveness, the UTAUT model accounts for 70% of the variability in users' technology usage intentions (Purwanto, 2020). In the study of technology adoption and human behavior, eight pivotal theories and models play crucial roles. These include the Motivational Model (MM), the Theory of Planned Behavior (TPB), and the Theory of Reasoned Action (TRA). Additionally, the Model of PC Utilization (MPCU) and the Innovation Diffusion Theory (IDT) significantly contribute to our understanding. The Combined TAM and TPB model (C-TAM-TPB) offers a comprehensive framework, while the Technology Acceptance Model (TAM) provides a well-established basis. Finally, the Social Cognitive Theory (SCT) is essential for understanding the adoption of technology.

This overview highlights the intricate interplay of these theories in shaping our understanding of technology adoption and human behavior (Venkatesh, 2016) (Jen, 2009).

Experiments have shown over and over that the UTAUT model is a good way to look at the factors that affect people's willingness to use technology and their plans to do so. This model surpasses its predecessors due to its comprehensive approach and robustness (Qasim, 2016). The UTAUT model's sensitivity to cultural variations makes it particularly suitable for international studies, effectively highlighting cultural differences and addressing translation challenges (Venugopala, 2016). Additional constructs, such as hedonic motivation (the fun and enjoyment derived from using technology), price value, and habit, have been integrated into the UTAUT framework (Brown, 2005) (Venkatesh, 2016). Research indicates that users are more likely to adopt a technology if they find it enjoyable. Habit can be defined in two ways: as past behavior or as the degree to which an individual performs behaviors automatically with smart technology due to learning. The latter definition aligns with information technology usage, with numerous studies confirming a strong correlation between previous behavior and technology acceptance.

While the UTAUT model encompasses most variables necessary for understanding technology acceptance and usage intentions, the relative significance of the four major constructs can fluctuate broadly and inconsistently, especially across different countries. This variability highlights the necessity for researchers to meticulously select the constructs they incorporate in their studies and to opt for suitable data analysis methods to guarantee valid results (Attuquayefio, 2014). Moreover, it might be essential to adjust the model to adequately account for variations among nations and specific objectives (Cheng, 2011). With EVs, it is vital to consider several key factors. The adoption of high-value technologies like EVs necessitates addressing traditional technology adoption concerns, such as the impacts of grid construction, incentive limitations, government support policies, and charging infrastructure interfaces, in addition to trust and network externalities (Venkatesh, 2016). The subsequent sections will scrutinize these crucial components within our proposed conceptual framework.

*Electric Vehicles (EVs):*
EVs, which include battery electric vehicles (BEVs) and plug-in hybrid electric vehicles (PHEVs), rely on electric drivetrains and batteries, offering a cleaner alternative to traditional combustion engines. While BEVs operate solely on electric power, PHEVs combine electric and gasoline systems, providing flexibility and efficiency as the transition to fully EVs progresses. There is a notable dearth of academic studies specifically addressing the challenges associated with owning an EV in Nigeria. New research using quota sampling and quantitative methods has shown that factors like high unemployment rates, limited access to energy, and limited finances have a big effect on people's desire to use sustainable energy technology. Nevertheless, the available literature has not extensively scrutinized crucial matters such as insufficient infrastructure, the absence of government assistance, and the need for facilitating conditions (Un-Noor, 2017).

Prior studies have explored the technical and economic viability of EV charging systems in Nigeria, with a specific focus on incorporating renewable energy sources. These studies have demonstrated that integrating renewable energy with EV infrastructure has the potential to significantly decrease carbon emissions, thereby facilitating the widespread adoption of EVs in Nigeria and throughout Africa (Ajao Q. a., 2023). Despite these observations, the extensive adoption of electric cars in Nigeria still faces challenges, primarily within the energy sector. This highlights the need for more comprehensive research on the attitudes and intentions of Nigerian residents, specifically regarding infrastructure, government aid, and the affordability of EVs. The study explores the factors driving Nigeria's EV adoption, focusing on infrastructure, government support, and charging facilities. It highlights the increasing popularity of EVs in the region, identifies barriers and EV adoption factors, and proposes future strategies for implementation.

*Facilitating Conditions:*

Refers to the degree to which an organization's technical and operational framework supports the utilization of a certain technology (Attuquayefio, 2014). These conditions are crucial for the effective adoption and integration of new technologies. They significantly influence technology adoption, often working alongside suitability. Facilitating conditions include three core constructs: perceived behavioral control, which covers internal constraints (skills and knowledge) and external constraints (resource availability and support); objective environmental factors that ease technology use, such as user support services, training programs, and necessary infrastructure; and compatibility, which measures how well the new technology aligns with users' values and needs (Venkatesh, 2016). In high-tech adoptions like EVs, facilitating conditions are particularly important. The availability of supporting infrastructure, such as charging stations and maintenance services, highlights how facilitating conditions play a part in predicting behavioral intentions to embrace new technology (Hsu, 2022). Facilitating conditions might include integration with existing transportation infrastructure, the availability of skilled maintenance services, and widespread charging stations. Government policies and incentives also play a significant role. Meeting these conditions can enhance users' willingness and ability to adopt EVs, making facilitating conditions a critical factor in the successful deployment of new automotive technologies (Gicha, 2024).

*Effort Expectancy:*

A core concept in the Unified Theory of Acceptance and Use of Technology (UTAUT) framework, gauges how easy and comfortable users find a technology. This concept includes factors such as the ease of learning to operate an EV, the simplicity of its user interface, and the accessibility of charging infrastructure. For new adopters and those contemplating the transition to EVs, these factors are critical. While their significance may diminish as users become accustomed to the technology, they are initially crucial for acceptance. A potent determinant of technology acceptance, effort Expectancy frequently exercises its most significant influence during the initial adoption stages (Qasim, 2016). The perceived ease or complexity of using a technology can heavily influence a user's willingness to adopt it (Venkatesh, 2016) (Ajao Q. a., 2023). Effort Expectancy and Performance Expectancy are indirectly connected when it comes to shaping a user's behavioral intentions. If users think that using a certain technology will require significant effort, they might start doubting its worthiness; in other words, their Performance Expectancy takes quite an expectancy hit. However, as time moves on, this link between Effort Expectancy and behavioral intentions tends to lose some steam. As users become more familiar with technology, the perceived effort to use it diminishes as a barrier (Dwivedi, 2011). This trend explains why some studies, such as one on e-recruitment systems, have not found Effort Expectancy to be a significant factor.

*Performance Expectancy:*

A key factor in embracing technological advancements involves an understanding of a technology's usefulness, potential benefits compared to other solutions, and the expected results it could yield. Defined in 2003 as the extent to which a user believes a specific technology will enhance their performance, Performance Expectancy is consistently identified as a robust predictor of technology acceptance (Venkatesh, 2016). Performance Expectancy reflects the belief that technology will effectively help users achieve their goals. With EVs, this includes improved travel efficiency, reduced environmental impact, and long-term cost savings. Without clear, tangible benefits, users are less likely to adopt the technology. Relative advantage pertains to the belief that a technology surpasses its predecessors or current alternatives. In the context of EVs, this could mean lower operating costs, enhanced driving experiences, and advanced safety features. Users are more inclined to adopt EVs if they perceive these vehicles as significantly superior to traditional internal combustion engine vehicles. Outcome expectations involve the anticipated positive results from using the technology. With EVs, these outcomes might include reduced emissions, contributions to sustainability, and alignment with personal or organizational values regarding environmental responsibility. By understanding and leveraging performance expectancy, stakeholders can more effectively promote EV adoption through targeted marketing and educational campaigns that emphasize efficiency, cost- effectiveness, and environmental benefits, thus enhancing perceived usefulness and relative advantage (Ajao Q. a., 2023).

*Social Influences:*

Also known as external influences, play a crucial role in how people decide to adopt new technology. This concept refers to the societal pressure to either engage in or avoid certain behaviors. Research shows that social influence strongly impacts behavioral intentions, shaping customer behavior through identification, internalization, and compliance.

Identification and internalization involve changing beliefs for potential status gains, while compliance is driven by social pressure (Venkatesh, 2016). People often look to their social circles for guidance to ease the anxiety of adopting innovations (Slade, 2015). This reliance on social networks underscores the importance of social influence in making informed decisions. Moreover, external influences and social image are strong predictors of consumer behavior. When it comes to EVs, social influence significantly affects adoption rates. Potential buyers often seek reassurance and advice from family, friends, and colleagues who already use EVs. Positive reviews and observed benefits can encourage adoption, while negative feedback or a lack of current users can hinder it. Recognizing the role of social influences is key for stakeholders promoting EV adoption. Effective strategies might include marketing campaigns with testimonials from early adopters, endorsements from influential figures, and community engagement initiatives. By addressing the social aspects of technology adoption, efforts to boost EV uptake can be more successful, leading to significant environmental and economic benefits (Malima, 2023).

*Trust:*

Trust in EVs is crucial for their acceptance and use. Factors affecting trust include safety, technical dependability, brand standing, data confidentiality, user satisfaction, governmental approvals, and community impact. Safety records, transparent information, reliable systems, robust cybersecurity measures, clear data use rules, strong data protection measures, user-friendly interfaces, intuitive design, and prompt customer service enhance trust (Venkatesh, 2016) (Malima, 2023) (Franke, 2015). Regulatory approvals, certifications, and testimonials from friends and family can further promote acceptance.

*Network Externalities:*

When more people start using a particular product or service, its worth tends to grow. This phenomenon is known as network externalities. This concept is pivotal in understanding the dynamics influencing the adoption of EVs. As more EVs are used, the demand for and consequently the investment in charging infrastructure grows, enhancing convenience and reducing range anxiety for users. This, in turn, makes EVs more attractive and feasible for potential owners (Ajao Q. a., 2023). A well-developed charging network not only improves accessibility but also contributes to economies of scale that lower the costs associated with EVs. Increased production leads to reduced costs for batteries, vehicle components, and charging infrastructure, making EVs more affordable and encouraging further adoption. Additionally, as more EVs are adopted, manufacturers are likely to increase investment in research and development, resulting in technological advancements that enhance vehicle performance, safety, and user experience. Moreover, a growing EV market stabilizes and possibly enhances the resale value of these vehicles, making them a more appealing investment (Malima, 2023) (Franke, 2015). A larger user base also ensures a robust secondary market and better availability of services such as maintenance and repairs, which further removes barriers to adoption. The increasing visibility of EVs boosts social acceptance, influencing more people to adopt this technology. Governments and regulatory bodies are also more likely to support and incentivize a widely adopted technology with favorable policies, further encouraging EV adoption. Thus, network externalities create a positive feedback loop that significantly propels the market growth of EVs.

**Research Methodology**

This research utilized the UTAUT model, renowned for its comprehensive explanatory capacity, derived from various established theories within the technology acceptance domain (Venkatesh, 2016). Originally used to investigate the factors influencing mobile payment behaviors, this model has been aptly adapted to assess the behavioral intentions towards EVs in Nigeria. Through an extensive literature review focused on EV adoption, we identified several key constructs that were not included in the original UTAUT model but have shown significant influence on acceptance. Simultaneously, some constructs previously deemed central were found to be less impactful or relevant for the acceptance of this sophisticated technology.

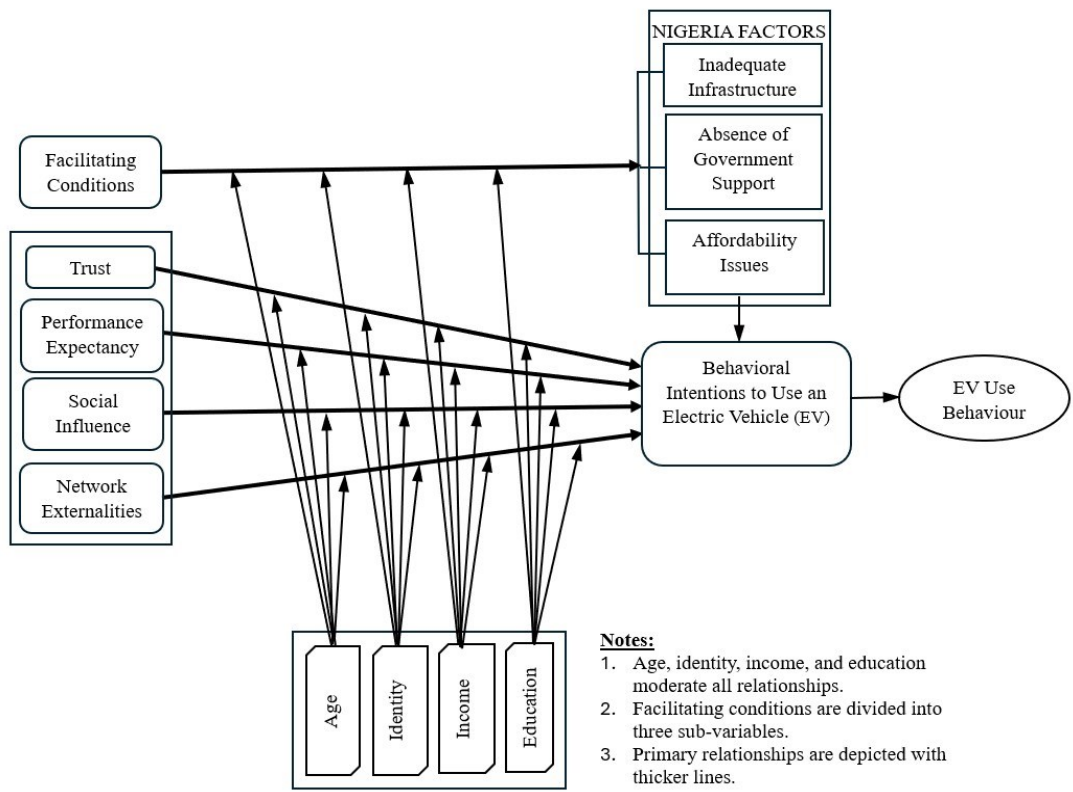

**Figure 1:** Proposed Research Model

This research incorporates specific challenges associated with EV adoption, such as the effects on power grid development, affordability constraints, government incentives, and integration with existing charging infrastructure. These factors, though non-traditional and often overlooked in sectors like auto-manufacturing where high user ratings imply reliability, are crucial. The study also looks at more common factors that affect how well EVs fit into Nigeria's society, economy, and infrastructure, such as trust, performance expectations, social influence, and network externalities. Figure 1 displays the updated UTAUT model, which focuses on the altered factors that could impact EV acceptance in Nigeria. This approach positions the research to offer crucial insights into the adoption dynamics of transformative technologies. Table 1 provides a detailed summary of the local Nigerian factors integrated into our enhanced conceptual framework.

**Table 1:** A Summary of Local Factors Used in the UTAUT Model

| Nigerian Local Variables | Details |
| --- | --- |
| Poor Infrastructure | 1. Unreliable Power Supply |
|  | 2. Few Charging Stations |
|  | 3. Substandard Roads |
| Affordability Issues | 1. Costly Purchase Price |
|  | 2. Home Charger Installation |
|  | 3. Higher Insurance Premiums |
| Governmental Support | 1. Tax Benefits and Subsidies |
|  | 2. EV Purchase Grants/Loans |
|  | 3. Regulatory Guidelines and Policies |

**Research Objective and Hypotheses**

This study investigates the elements that shape the acceptance of EVs in Nigeria by reviewing existing literature to pinpoint key factors that might influence consumer attitudes toward technology acceptance. We develop several hypotheses using a designed research model to test these influences. A crucial aspect under consideration is the role of facilitating conditions (FC), which gauge a nation's capability to embrace and integrate new technologies effectively. For instance, if the local population views environmentally friendly solutions such as EVs as impractical due to current infrastructure constraints, their willingness to adopt these technologies may decrease. This supports our hypothesis that the perception of sustainability significantly increases the likelihood of its adoption, highlighting the critical role of facilitating conditions in technology adoption. Thus, we hypothesize that:

**H1:** Creating facilitating conditions **(FC)** helps to influence people to utilize an EV.
Trust denotes the readiness of a consumer to risk vulnerability based on the vendor's traits, and with expectations that they will deliver promised services. Similarly in the EV industry, for smooth operation and fulfillment of promises by manufacturers, it is imperative to have faith in their reliability technologically as well as functionally. Consequently

**H2:** Trust **(T)** significantly influences behavioral intentions **(BI)** to purchase an EV.

Performance expectancy refers to the idea of how much someone or people think that using an EV in their daily living will boost the overall standard and enjoyment of their life. Consumers are more likely to embrace EVs if they perceive these vehicles as cost-effective or capable of improving transportation efficiency. Previous studies have shown that positive perceptions of the benefits associated with EVs significantly impact their adoption (Al-Saedi, 2020). Therefore, in Nigeria, consumers' behavioral intentions (BI) to use EVs are positively affected by performance expectancy (PE). Thus:

**H3:** The optimistic prospect of performance expectancy **(PE)** can effectively sway behavioral intention **(BI)** toward owning an EV.

Social influence pertains to the external persuasion experienced by individuals within their immediate social circles when it comes to embracing or disregarding novel technology. In Nigeria, such societal sway is particularly potent in influencing individual choices regarding new technological adoptions, as collective viewpoints can considerably steer personal decisions (Venkatesh, 2016). Therefore:

**H4:** Social influence **(SI)** significantly boosts the likelihood of behavioral intentions **(BI)** that people decide to purchase an EV.

The concept of network externalities suggests that the appeal and utility of a product rise with an increase in its users. This theory is clearly illustrated through EV usage as the number of users and the network of charging stations grow, the vehicles become more useful and attractive (Haruvy, 1998). Therefore, we propose the following hypothesis:

**H5:** Network externalities positively impact the behavioral intentions **(BI)** to adopt an EV.

**Table 2:** Instrument Design and Creation

| MAIN QUESTIONNAIRES | |
|---|---|
| **Facilitating Conditions** | **Trust** |
| I am concerned about EV affordability compared to ICEVs. | I highly trust the reliability of EVs. |
| I find the scarcity of sufficient charging facilities concerning. | I'm confident in the security of the data my EV collects. |
| I care about the integration with Renewable Energy Sources. | My EV promptly notifies me of any issues. |
| I care about the development and reliability of the electric grid. | I trust the systems and technology of EVs. |
| The unreliability of the electricity supply is a concern for me. | |
| | |
| **Performance expectancy** | **Network externalities** |
| I perceive significant daily benefits from driving an EV. | I think more EV users will lead to: |
| EVs facilitate more efficient transportation for me. | Reduced maintenance costs for all EV users. |
| Using an EV boosts my daily productivity. | An expansion in the availability of service centers. |
| Driving an EV helps me accomplish tasks more effectively. | Improvement of EV charging infrastructure and services. |
| | |
| **Social influence** | **Behavioral intentions** |
| People important to me advised me that I should drive an EV. | I plan to acquire an EV eventually. |
| Influential individuals in my life suggested I purchase an EV. | I aim to use an EV frequently for my transportation needs. |
| Those who influence my decisions support my use of an EV. | My daily transportation will likely include regular EV use. |

**Instrument Design and Creation**

In this study, we systematically analyzed key components of our enhanced model to assess EV adoption. We crafted specialized survey questions, inspired by prior research, to gauge the accuracy of our instrument. Respondents rated their views and intentions regarding EV usage on a scale from "strongly support" to "strongly oppose." To accommodate non-English speakers, we translated the survey items, which are listed in Table 1. We utilized a back-translation method, translating the questions into another language and then translating them back into English to ensure the translations were accurate and culturally appropriate (Brislin, 1970). This process helped maintain linguistic accuracy and contextual relevance for our Nigerian participants, ensuring that the survey items resonated culturally.

Each step of this process was designed to enhance the clarity and trustworthiness of the data collected, ultimately providing insights into the factors influencing the acceptance of EVs in diverse settings (Haruvy, 1998). The final version of the questionnaire, refined through this rigorous translation and adaptation process, was then disseminated to participants. This approach ensured the survey items were both understandable and relevant to the local context, thereby enhancing the reliability of the data collected on the adoption factors influencing EV acceptance.

**Table 3:** Sampled Data Demographics

| Identity | Tally (%) | Age | Tally | Income (₦) | Tally | Education | Tally |
|---|---|---|---|---|---|---|---|
| Female | 13,590 (33.76) | Below 25 | 8,550 | < ₦50,000 | 8,950 | College | 2,850 |
| Male | 22,300 (55.38) | 25 - 45 | 14,890 | ₦50,000 - ₦100,000 | 12,150 | Polytechnic | 8,050 |
| Unknown* | 4,360 (10.83) | Above 45 | 16,260 | > ₦100,000 | 16,000 | University | 16,800 |
| | | Unknown* | 550 | | | Vocational | 7,750 |
| | | | | | | Unknown* | 4,800 |
| **TOTAL** | 40,250 (100) | 40,250 | | | 40,250 | 40,250 | |

*Unknown = Not Specified

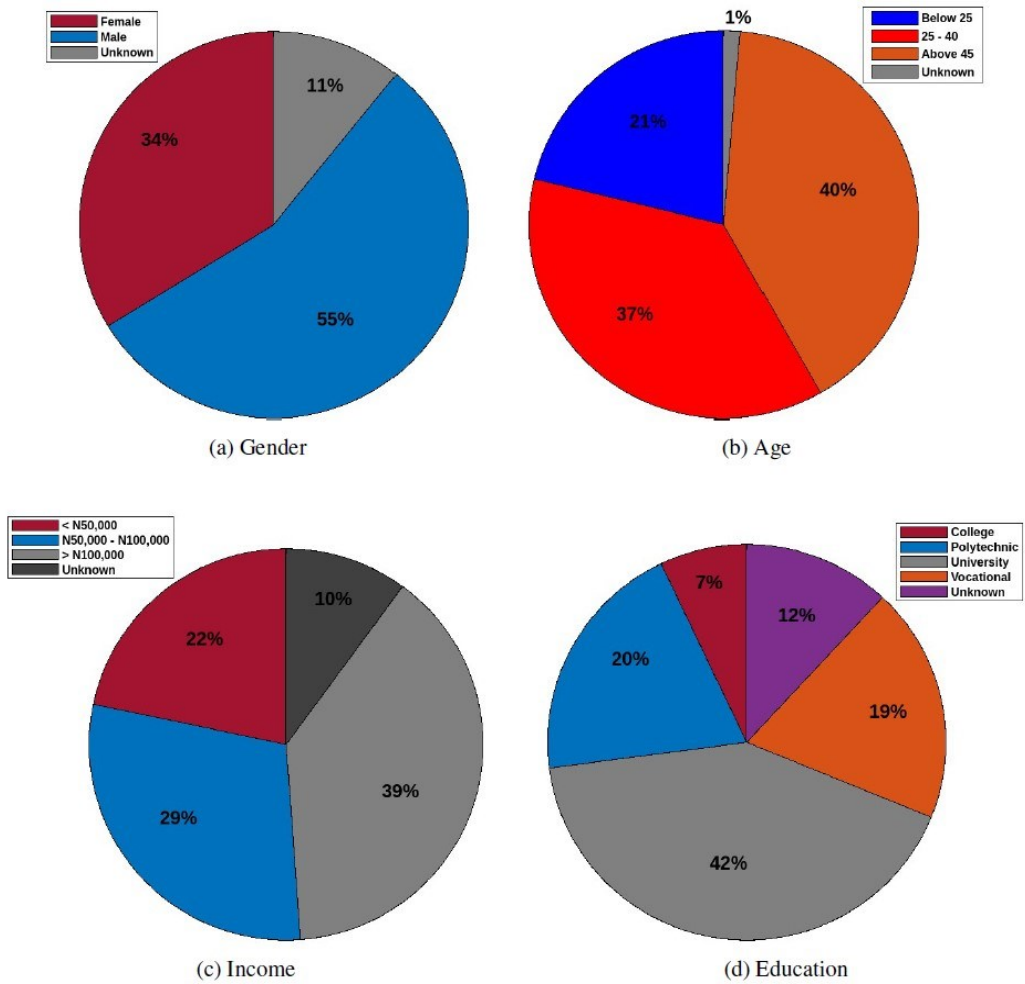

Figure 2: Demographic distribution data

**Sampling Technique and Questionnaire**

Due to the scarcity of EV ownership in Nigeria, assembling a target group for this study presented quite a challenge. Our research team attempted to reach 50,000 people by personally handing out many of the surveys. To expedite the collection of information from respondents, we enlisted the aid of ten willing volunteers, including secondary school teachers, university students, and lecturers. We further broadened our survey pool with digital distributions among Nigerians residing overseas, attempting to increase the representational diversity within the responses gathered. Respondents weren't just randomly selected; they participated in preliminary discussions on EVs before receiving survey forms, ensuring they understood the queries, and thereby providing more accurate feedback possibilities. The survey was conducted between January and February 2024, allowing participants to join willingly without any incentives being offered. Although we aimed for a sample size of 50,000, we managed to distribute 45,527 surveys, successfully recovering 40,255. Following a preliminary visual assessment, 40,250 surveys were deemed usable. This research examined a broad cross-section of Nigerians, encompassing a range of ages, education levels, and economic statuses. While it did not precisely reflect the region's demographics, it nonetheless provided a comprehensive overview for assessing how well EVs might be embraced nationwide.

Table 4: Collinearity Data Analysis

| Enhanced Variable | Tolerance | Variance Inflation Factor (VIF) |
| --- | --- | --- |
| Facilitating Conditions (FC) | 0.511 | 1.788 |
| Trust (T) | 0.625 | 1.490 |
| Performance Expectancy (PE) | 0.522 | 1.899 |
| Network Externalities (NE) | 0.570 | 1.801 |
| Social Influence (SI) | 0.750 | 1.509 |

**Data Evaluation and Interpretation**

40,250 surveys were gathered and arranged in Excel before being loaded into SPSS and AMOS 20 software. The data breakdown by gender, age, education, and income is shown in Figure 2. SPSS was used for initial data grouping and simple stats. For more detailed analyses and testing relationships, AMOS was utilized. All required analyses were completed to answer research questions and validate hypotheses (Collett, 2021). The data analysis procedure will be detailed, and outcomes closely reviewed in upcoming sections.

**Preliminary Assessment**

To assess the impact of outliers on a specialized model for evaluating EV adoption in Nigeria, we conducted a multiple regression analysis with 40,255 responses. We identified and excluded redundant outliers with standard residuals during case-wise diagnostics, thereby reducing the effective sample size to 40,250. The demographic data summarized in Table 2 showed a predominance of male respondents, mostly adults with university degrees and falling into the Nigerian high-income bracket, indicating a population that is relatively knowledgeable about EVs. The study initially screened for extreme correlations among variables, which could distort the regression results. To manage potential multicollinearity, we conducted collinearity tests. We will discuss further details later. Table 4 details the tolerance and variance inflation factor (VIF) metrics used to evaluate correlated data. Tolerance measures the predictive redundancy of variables, while VIF evaluates the extent to which predictor correlation inflates the variance of a regression coefficient. Values below 0.1 for tolerance or above 2.5 for VIF typically indicate multicollinearity (Venkatesh, 2016) (Abu-Shanab, 2009) (Hair, 2012). However, all observed values were within acceptable limits, indicating no significant multicollinearity issues. We also assessed internal consistency using Cronbach's alpha, with acceptable levels noted above 0.75. Results, shown in Table 5, (i.e., Cronbach's Alpha ($C'A$) with the variables, means, and computed standard deviations (SD)) confirmed high consistency among survey items, supporting the reliability of our instruments in evaluating EV adoption factors in Nigeria.

Table 5: Internal Consistency Measurement

| Variables | Item | Mean | (SD) | (C'A) | No. of Items |
|---|---|---|---|---|---|
| Facilitating Conditions (FC) | 1 | 3.46 | 1.12 | 0.896 | 4 |
|  | 2 | 3.54 | 0.78 |  |  |
|  | 3 | 3.59 | 1.00 |  |  |
|  | 4 | 3.68 | 1.00 |  |  |
| Trust (T) | 5 | 2.96 | 1.17 | 0.722 | 4 |
|  | 6 | 3.02 | 1.01 |  |  |
|  | 7 | 3.20 | 1.01 |  |  |
|  | 8 | 3.20 | 1.01 |  |  |
| Performance Expectancy (PE) | 9 | 3.51 | 1.01 | 0.813 | 4 |
|  | 10 | 3.64 | 0.88 |  |  |
|  | 11 | 3.83 | 0.85 |  |  |
|  | 12 | 3.40 | 1.00 |  |  |
| Network Externalities (NE) | 13 | 3.32 | 1.21 | 0.802 | 3 |
|  | 14 | 3.63 | 0.88 |  |  |
|  | 15 | 3.55 | 1.00 |  |  |
| Social Influence (SI) | 16 | 3.05 | 0.99 | 0.759 | 3 |
|  | 17 | 3.11 | 0.84 |  |  |
|  | 18 | 3.20 | 1.00 |  |  |
| Behavioral Intention (BI) | 19 | 3.14 | 1.04 | 0.804 | 3 |
|  | 20 | 3.38 | 0.88 |  |  |
|  | 21 | 3.28 | 1.09 |  |  |

**Factor Analysis**

We performed a confirmatory factor analysis to scrutinize how measurable variables represent defined factors in the adoption of EV. This study, which is shown in Table 6, uses measurements like factor loadings, construct reliability, average variance extracted (AVE), maximum shared variance (MSV), and average shared variance (ASV) to check how well our model works and whether the instruments we made are right for use in the area. We consider factor loadings above 0.5 to be significant; all factors in this study met this standard. Construct reliability (CR) measures each construct's internal consistency, with values above 0.7 denoting excellent reliability Hair (2009). In our study, all constructs recorded CR values exceeding this level, values over 0.5 indicate satisfactory convergence, indicating that the construct successfully captures more variance than errors. All the constructs in our analysis met this requirement. Our findings support convergent validity, which validates that factors within a construct distinctly measure various dimensions without significant overlap when both CR and AVE values exceed 0.5. When both the ASV and MSV are below the AVE, we achieve discriminant validity, which confirms that constructs are distinct and not overly similar. Our model also met this condition, confirming the distinctiveness of constructs relevant to EV adoption in Nigeria (Hair, 2012).

**Table 6:** Factor Analysis

| Item | Instruments | Factor Loading | CR* | AVE* | MSV* | ASV* |
|---|---|---|---|---|---|---|
| **Facilitating Conditions (FC)** | Question 1 | 0.81 | 0.847 | 0.659 | 0.360 | 0.288 |
| | Question 2 | 0.75 | | | | |
| | Question 3 | 0.82 | | | | |
| | Question 4 | 0.87 | | | | |
| **Trust (T)** | Question 5 | 0.99 | 0.897 | 0.690 | 0.423 | 0.343 |
| | Question 6 | 0.76 | | | | |
| | Question 7 | 0.74 | | | | |
| | Question 8 | 0.81 | | | | |
| **Performance Expectancy (PE)** | Question 9 | 0.77 | 0.813 | 0.521 | 0.490 | 0.386 |
| | Question 10 | 0.76 | | | | |
| | Question 11 | 0.66 | | | | |
| | Question 12 | 0.69 | | | | |
| **Network Externalities (NE)** | Question 13 | 0.84 | 0.863 | 0.690 | 0.490 | 0.404 |
| | Question 14 | 0.91 | | | | |
| | Question 15 | 0.71 | | | | |
| **Social Influence (SI)** | Question 16 | 0.79 | 0.819 | 0.603 | 0.303 | 0.260 |
| | Question 17 | 0.68 | | | | |
| | Question 18 | 0.85 | | | | |
| **Behavioral Intention (BI)** | Question 19 | 0.82 | 0.866 | 0.681 | 0.490 | 0.374 |
| | Question 20 | 0.87 | | | | |
| | Question 21 | 0.79 | | | | |

**Table 7:** The set of relationships and their corresponding standardized betas

| Hypotheses # | Relationship | Result | Stand. Beta |
|---|---|---|---|
| H1 | Facilitating conditions (FC) → Behavioral intentions (BI) | Supported | 0.44 |
| H2 | Trust (T) → Behavioral intentions (BI) | Supported | 0.37 |
| H3 | Performance Expectancy (PE) → Behavioral intentions (BI) | Supported | 0.25 |
| H4 | Network externalities (NE) → Behavioral intentions (BI) | Supported | 0.20 |
| H5 | Social Influence (SI) → Behavioral intentions (BI) | Supported | 0.15 |

**Hypotheses Validation & Model Fit Criteria**

Before evaluating our hypotheses related to the acceptance of EVs in Nigeria, it is essential to scrutinize the foundational assumptions about our variables. The bivariate relationships among all variables suggest multidimensionality. As detailed in Table 5, the correlation matrix elucidates how the independent variables; trust, effort expectancy, performance expectancy, network externalities, and social influence significantly correlate with behavioral intentions, each showing a significance level of $p < 0.01$. Notably, performance expectancy and network externalities exhibit the most substantial bivariate effects on behavioral intentions. Further analysis of the correlation matrix reveals the absence of extreme values (neither too low nor excessively high), which could indicate insignificant relationships or redundancy among measures, respectively. Specifically, no correlation coefficients are below 0.2 or exceed 0.8, supporting our theoretical assumption that each independent variable significantly relates to behavioral intentions. To assess the model's fit, various metrics were used—chi-square, degrees of freedom, the Comparative Fit Index (CFI), and the Root Mean Square Error of Approximation (RMSEA)—through the AMOS 20 software. The values presented in Table 6, fall within acceptable limits, suggesting a good fit for our structured research model.

The predictive results from the model reveal that all independent variables significantly influence behavioral intentions, except for effort expectancy. The model explains 58% of the variance in behavioral intentions ($R^2 = 0.579$, adjusted $R^2 = 0.567$), which is considered substantial explanatory power—surpassing the threshold of 0.4 as suggested by (Cohen, 1983). Table 7 details the estimated relationships, standardized beta coefficients, and the outcomes of hypothesis testing.

The predictive equation for behavioral intentions (BI) about EVs in Nigeria can be summarized as follows:

$$BI = 0.44 \, FC + 0.17 \, T + 0.25 \, PE + 0.2 \, NE + 0.15 \, SI + error \qquad (1)$$

This refined analysis helps clarify the significant drivers of EV acceptance in Nigeria, underpinning the need for targeted strategies to enhance acceptance rates.

**Table 8:** Model Fit Measures

| Index | Value | Threshold values |
|---|---|---|
| Chi-square | 333.092, $P < 0.001$ | Significant |
| Degree of freedom | 155 | |
| $\chi^2/df$ (deg. of freedom) | 2.210 | < 0.3 |
| CFI | 0.827 | > 0.90 |
| TLI | 0.808 | > 0.90 |
| RMSEA | 0.059 | X* = Significant if <0.07, with a sample size >40,250 and CFI >0.827 |

*T\* = Threshold Values. 2\* = Degree of Freedom.*

## Results and Discussions

As highlighted in Table 7, **facilitating conditions (H1)** emerged as the most significant predictor of EV adoption in Nigeria, underscoring the critical role of robust infrastructure such as charging stations and reliable energy supply. Additionally, government support through subsidies, tax incentives, and supportive policies is essential in addressing concerns about the practicality and affordability of EVs, particularly in metropolitan areas where competitive pricing and governmental backing can substantially enhance their appeal. More pressing factors often overshadow the potential of green technology in Nigeria. This observation fits with earlier research that found that people are much more likely to use new technologies when conditions are very good for them (Qasim, 2016) (Venkatesh, 2016).

The inclusion of **trust (H2)** and facilitating conditions enriches the model. Trust is vital, reflecting the necessity for user confidence in both the technology and manufacturers, given the financial and safety considerations associated with EVs.

**Performance expectancy (H3)** emerged as a significant predictor of acceptance, indicating that users perceive EVs as offering substantial benefits over traditional vehicles. The evidence corroborates the concept that users are likely to accept new technologies when they see clear advantages over existing solutions (Hsu, 2022) (Slade, 2015).

Similarly, **social influence (H4)** significantly impacts behavioral intentions, confirming that the opinions of peers and the community significantly influence the choice to possess EVs. This underscores the role of societal norms and peer perceptions in tech adoption, as highlighted in earlier studies (Venkatesh, 2016).

**Network externalities (H5)** proved to be the most influential predictor, indicating that the likelihood of EV acceptance increases as more charging stations become available and as more people use EVs. This effect suggests that achieving a critical mass of users and infrastructure is pivotal for widespread acceptance. This finding supports theories that network externalities are crucial in driving technology acceptance, particularly when the technology facilitates essential daily functions like transportation. This study underscores the complex interplay of user expectations, societal influences, and practical considerations in the acceptance of EVs in Nigeria, offering valuable insights for policymakers, manufacturers, and stakeholders aiming to enhance the penetration of sustainable transport solutions in the region. This study on the adoption of EVs in Nigeria has uncovered several critical factors affecting their acceptance. Our findings offer valuable insights for both researchers and practitioners in the field (Ajao Q. a., 2023).

## Conclusion and Future Work

This study identifies five key factors influencing the decision to purchase an EV in Nigeria: facilitating conditions (FC), trust (T), social influence (SI), performance expectancy (PE), and network externalities (NE). Bivariate correlations confirm direct predictive associations, which supports our hypotheses. When we added these predictors to our improved model, which looks at important factors when conditions are favorable like poor infrastructure, lack of government support, and problems with affordability, they had a big effect because they were connected and overlapping. The results reveal that all examined factors except for effort expectancy significantly impact behavioral intentions, corroborating all our hypotheses. Notably, facilitating conditions emerged as the most influential predictor of EV adoption in Nigeria, underscoring their critical role. These determinants collectively accounted for 60% of the variance in behavioral intentions, underscoring their importance in understanding EV adoption.

**Considerations for Researchers**

This research highlights the key factors that influence the acceptance of EVs in Nigeria, focusing on support systems (i.e., facilitating conditions), trust, social influence, performance expectations, and network effects. Although our model did not directly address ease of use, previous studies have shown that it may do so through performance expectations, suggesting a potential area for further research. While traditional models of technology acceptance do not always consider facilitating conditions, they were found to be very important in the adoption of EVs, which could make these models better at predicting the future. Understanding these factors is crucial for industry stakeholders and policymakers in Nigeria as they develop strategies to increase EV uptake. Well-developed support systems, such as extensive charging networks, consistent power supply, and reliable maintenance services, play a pivotal role in influencing consumer decisions toward EV adoption. Strengthening the national grid and integrating sustainable energy solutions are also key to encouraging widespread EV use, supporting sustainability, and reducing environmental impacts. The government needs to help with this transition by offering incentives, tax breaks, and policies that remove practical and financial barriers. This is especially true in cities, where these factors make EVs much more appealing.

**Considerations for Practitioners**

Understanding the factors that influence the acceptance of EVs is critical for businesses, including automotive manufacturers and service providers in Nigeria. The significance of network externalities suggests that achieving a critical mass of users and ensuring service availability are essential. Companies should prioritize the development of robust networks of charging stations to mitigate range anxiety and enhance user convenience. Additionally, providing widespread and accessible maintenance services is necessary to support EV owners and improve reliability. Our model helps automakers formulate effective marketing strategies for potential EV customers. The significant attention respondents gave to network externalities implies that achieving a critical mass of users and service centers is paramount. Customers are more inclined to adopt EVs if there are sufficient service centers and a large user base for such technology. Establishing trust is also vital. Building a trustworthy brand entail communicating transparently about the benefits and reliability of EVs, addressing concerns, and showcasing proven performance and safety records.

Cost and affordability are significant considerations in Nigeria, where price sensitivity is high. Companies must work on making EVs more affordable through competitive pricing strategies, financing options, and government incentives. Grid infrastructure development is another critical factor. Ensuring a stable and efficient electricity supply is vital for the extensive endorsement of EVs. This includes upgrading existing grid infrastructure to support the additional load from EV charging stations. Integrating renewable energy sources into the grid could enhance the appeal of EVs by promoting sustainability and reducing environmental impact. By focusing on these areas: network externalities, trust, cost, affordability, grid infrastructural development, and the integration of renewable energy—automakers can encourage the adoption of EVs in Nigeria. This approach will ultimately enhance the sustainability and efficiency of the regional transportation system.

**Technical Recommendations**

To enhance EV adoption in Nigeria, a coordinated effort involving automakers, government policymakers, and other key stakeholders is essential, with a particular focus on network externalities and facilitating conditions. Network externalities emphasize the increasing value of EV ownership as more people adopt the technology. Government policies that incentivize EV purchases, such as subsidies, tax breaks, loans, and investments in public charging infrastructure, can accelerate this effect. As the EV user base expands, it will create a positive feedback loop, encouraging further adoption and fostering the growth of supportive infrastructure. Facilitating conditions, which include the necessary support systems and infrastructure for EVs, are critical to their widespread adoption. Governments play a pivotal role in this area by developing and implementing policies that ensure a reliable power supply, expand charging networks, and promote renewable energy sources to power these vehicles. Collaborating with private sector stakeholders, including automakers and energy companies, to build and maintain this infrastructure is vital. Additionally, governments should focus on creating regulations that support EV development, such as mandating the inclusion of EV-ready infrastructure in new developments and setting standards for vehicle emissions and efficiency. At the same time, government policies can support these efforts by funding research and development initiatives that drive innovation in battery technology and vehicle performance.

Competitive pricing remains a significant factor in making EVs an attractive alternative to internal combustion engine vehicles (ICEVs). Governments can assist by offering financial incentives that reduce the initial cost of EVs, while automakers should consider working with stakeholders to improve and expand the charging infrastructure, ensuring that charging stations are widely accessible and conveniently located. Building trust in EV technology is also essential for fostering adoption. Trust plays a significant role in the acceptance of new technology, and system transparency is key to building this trust. Governments can enforce regulations that protect consumer data and ensure the reliability of EV systems, while automakers should ensure transparency by clearly communicating the storage and use of personal data. Consistency in quality across all models and regions is vital, as it assures consumers of the same high standards of performance and reliability worldwide. Through a collaborative effort involving automakers, government policymakers, and stakeholders, we can focus on these areas—network externalities, facilitating conditions, and trust—to create a more compelling value proposition for potential EV buyers in Sub-Saharan Africa. This comprehensive approach will contribute to a more sustainable and efficient transportation system in the region, ultimately driving the widespread adoption of EVs.

**Prospects for Further Studies**

This study faced limitations due to a small sample size compared to the overall population of the country in question, particularly in some regions of Nigeria, which restricted the generalizability of the findings regarding electric vehicle (EV) adoption. Generally, researchers prefer larger samples to mitigate biases and enhance the robustness of the research findings. Additionally, the validity and reliability of the designed instruments posed challenges to other researchers, especially when language variations affected the responses. Previous research has shown that the language used in a survey can have a big effect on the results. This shows how important it is to carefully think about and validate the questionnaire many times in future research. The selection of survey items to measure constructs was another area of concern. This study used items from existing research that could potentially face criticism from other researchers, especially when it comes to constructs such as facilitating conditions. Previous research indicated a limited impact on behavioral intentions, so we did not regularly scrutinize facilitating conditions as a factor in the context of technology adoption. This calls for a reevaluation of the items representing each construct to ensure they accurately reflect the intended measurements and improve content validity.

Furthermore, the theoretical framework could influence the choice of our enhanced model and selected variables (both traditional and non-traditional), suggesting that different variables may be suitable depending on the specific context. It is also important to look at how future models like the UTAUT-2 can be used to get more people in Nigeria to buy electric vehicles (Macedo, 2017) (Abu-Shanab E. a., 2013). These models consider things like price, habitual use, and hedonic motivation. Future research should focus on the aspects with the addition of demographic influences, particularly economic factors, which could play a crucial role in the financial decisions surrounding the adoption of EV technologies in Nigeria.

**Declarations:** The authors have no conflicts of interest to declare that are relevant to the content of this article.

**Acknowledgments:** We extend our gratitude to the various facilitators who volunteered to assist with the data collection process. Special thanks are due to the ten individuals—including high school teachers and undergraduate and master's students—who generously contributed their time and effort. We wish to clarify that the views expressed in this paper are solely those of the authors.

**Data Availability:** The data that support the findings of this study are available from the corresponding author upon reasonable request.

## Author Contributions

**Qasim Ajao:** Conceptualization (equal); Formal analysis (equal); Methodology (equal); Software (equal); Visualization (equal); Writing– original draft (lead). Nicholas Hamilton: Conceptualization (equal); Methodology (equal); Project administration (equal); Resources (equal); Supervision (equal); Writing – review & editing (equal).

**Lanre Sadeeq:** Conceptualization (equal); Methodology (equal); Project administration (equal); Supervision (equal); Writing – review & editing (equal). Patrick Moriarty: Funding acquisition (lead); Project administration (equal); Resources (equal); Supervision (equal); Writing – review & editing (equal).

## References

Abu-Shanab, E. a. (2009). Internet banking in Jordan: An Arabic instrument validation process. *Int. Arab J. Inf. Technology*.

Abu-Shanab, E. a. (2013). The influence of language on research results. *Management Research \& Practice*.

Ajao, Q. a. (2023). Drivers of Mobile Payment Acceptance: The Impact of Network Externalities in Nigeria. *arXiv preprint arXiv:2305.15436*.

Ajao, Q. a. (2023). Overview Analysis of Recent Developments on Self-Driving Electric Vehicles. *arXiv preprint arXiv:2307.00016*.

Aldhanhani, T. a. (2024). Future trends in smart green iov: Vehicle-to-everything in the era of electric vehicles. *IEEE Open Journal of Vehicular Technology*.

Al-Saedi, K. a.-E. (2020). Developing a general extended UTAUT model for M-payment adoption. *Technology in society*.

Attuquayefio, S. a. (2014). Using the UTAUT model to analyze students' ICT adoption. *International Journal of Education and Development using ICT*.

Breschi, V. a. (2022). Fostering the mass adoption of electric vehicles: A network-based approach. *IEEE Transactions on Control of Network Systems*.

Brislin, R. W. (1970). Back-translation for cross-cultural research. *Journal of cross-cultural psychology*.

Brown, S. A. (2005). Model of adoption of technology in households: A baseline model test and extension incorporating household life cycle. *MIS quarterly*.

Cheng, Y. a.-B.-K.-Y.-L.-H.-M.-M.-L. (2011). Mass absorption efficiency of elemental carbon and water-soluble organic carbon in Beijing, China. *Atmospheric Chemistry and Physics*.

Cohen, A. (1983). Comparing regression coefficients across subsamples: A study of the statistical test. *Sociological Methods and Research*.

Collett, K. A. (2021). Can electric vehicles be good for Sub-Saharan Africa? *Energy Strategy Reviews*.

Dwivedi, Y. K. (2011). *Governance and Sustainability in Information Systems. Managing the Transfer and Diffusion of IT: IFIP WG 8.6 International Working Conference, Hamburg, Germany, September 22-24, 2011. Proceedings.* Springer.

Franke, T. a. (2015). Advancing electric vehicle range displays for enhanced user experience: the relevance of trust and adaptability. *Proceedings of the 7th international conference on automotive user interfaces and interactive vehicular applications*.

Gicha, B. B. (2024). The electric vehicle revolution in Sub-Saharan Africa: Trends, challenges, and opportunities. *Energy Strategy Reviews*.

Hair, J. F. (2012). Multivariate data analysis. *Multivariate data analysis*.

Haruvy, E. a. (1998). Optimal product strategies in the presence of network externalities. *Information Economics and Policy*.

Hsu, C.-W. a.-C. (2022). What drives older adults' use of mobile registration apps in Taiwan? An investigation using the extended UTAUT model. *Informatics for Health and Social Care*.

Jen, W. a.-T. (2009). An integrated analysis of technology acceptance behaviour models: Comparison of three major models. *MIS REVIEW: An International Journal*.

Macedo, I. M. (2017). Predicting the acceptance and use of information and communication technology by older adults: An empirical examination of the revised UTAUT2. *Computers in human behavior*.

Malima, G. C. (2023). Are electric vehicles economically viable in sub-Saharan Africa? The total cost of ownership of internal combustion engine and electric vehicles in Tanzania. *Transport Policy*.

Purwanto, E. a. (2020). The intention and use behaviour of the mobile banking system in Indonesia: UTAUT Model. *Technology Reports of Kansai University*.

Qasim, H. a.-S. (2016). Drivers of mobile payment acceptance: The impact of network externalities. *Information Systems Frontiers*.

Slade, E. L. (2015). Modeling consumers' adoption intentions of remote mobile payments in the United Kingdom: extending UTAUT with innovativeness, risk, and trust. *Psychology and marketing*.

Un-Noor, F. a.-P. (2017). A comprehensive study of key electric vehicle (EV) components, technologies, challenges, impacts, and future direction of development. *Energies*.

Venkatesh, V. a. (2016). Unified theory of acceptance and use of technology: A synthesis and the road ahead. *Journal of the association for Information Systems*.

Venugopala, P. a. (2016). User Acceptance of Electronic Health Records: Cross Validation of Utaut Model. *Global Management Review*.